\documentclass[aps,prb,showpacs,twocolumn,floats]{revtex4}
\usepackage{graphicx}
\usepackage{subfigure}
\usepackage{epsfig}
\usepackage{dcolumn}
\usepackage{bm}
\usepackage[ansinew]{inputenc}
\usepackage{amsmath}
\usepackage{amsthm}
\usepackage[T1]{fontenc}
\usepackage{amssymb}
\usepackage{amsfonts}
\usepackage[english]{babel}
\usepackage{enumitem}

\begin{document}

\title{Narrow Zero-Field Tunneling Resonance in Triclinic Mn$_{12}$ Acetate Ribbons}
\date{\today}
\author{ I. Imaz$^1$,  J. Espin$^1$, D. Maspoch$^{1,2}$}
\affiliation{$^1$Institut Catal\`{a} de Nanotecnologia, ICN2, Esfera Universitat Aut\'{o}noma Barcelona (UAB), Campus UAB, 08193 Bellaterra,  Spain\\$^2$Instituci\'{o} Catalana de Recerca i Estudis Avan\c{c}ats (ICREA), 08100 Barcelona, Spain}
\author{J. Tejada, R. Zarzuela, N. Statuto}
\affiliation{Departament de F\'{i}sica Fonamental, Facultat de F\'{i}sica, Universitat de Barcelona, Mart\'{i} i Franqu\`{e}s 1, 08028 Barcelona, Spain}
\author{E. M. Chudnovsky}
\affiliation{Physics Department, Lehman College,
The City University of New York, 250 Bedford Park Boulevard West, Bronx, NY 10468-1589, USA}

\begin{abstract}
We report synthesis, structural characterization, and magnetic measurements of amorphous Mn$_{12}$-acetate ribbons with a triclinic short-range crystal order not previously seen in experiment. The ribbons exhibit the same structure of Mn$_{12}$ molecules and the same positions of tunneling resonances on the magnetic field as a conventional tetragonal Mn$_{12}$-acetate crystal. However, the width of the zero-field resonance is by at least one order of magnitude smaller, indicating very small inhomogeneous broadening due to dipolar and nuclear fields. Possible origins of this effect are discussed. 
\end{abstract}


\maketitle

\section{Introduction}\label{introduction}
Chemistry and physics of molecular magnets have been intensively studied in the last two decades \cite{Springer} after it was found that they provide an ultimate limit of the miniaturization of a permanent magnet \cite{Sessoli} and, on top of it, exhibit quantum tunneling of the magnetic moment \cite{Friedman-PRL1996,Hernandez-EPL1996,Barbara-Nature1996,MQT-book}. Other fascinating quantum effects observed in molecular magnets include quantum topological Berry phase \cite{WS}, magnetic deflagration \cite{CCNY-PRL2005,UB-PRL2005}, and Rabi oscillations \cite{Schledel-PRL2008,Bertaina-Nature2008}. Most recently, experiments with individual magnetic molecules bridged between conducting leads and molecules grafted on carbon nanotubes have been performed that permit readout of quantum states of individual atomic nuclei \cite{Wern-NatureNano2013,Wern-ASC-Nano2013}. Quantum superposition of spin states in magnetic molecules makes them candidates for qubits -- elements of quantum computers \cite{Nanotech-2001}. 

\begin{figure}[htbp!]
\includegraphics[width=9.0cm,angle=0]{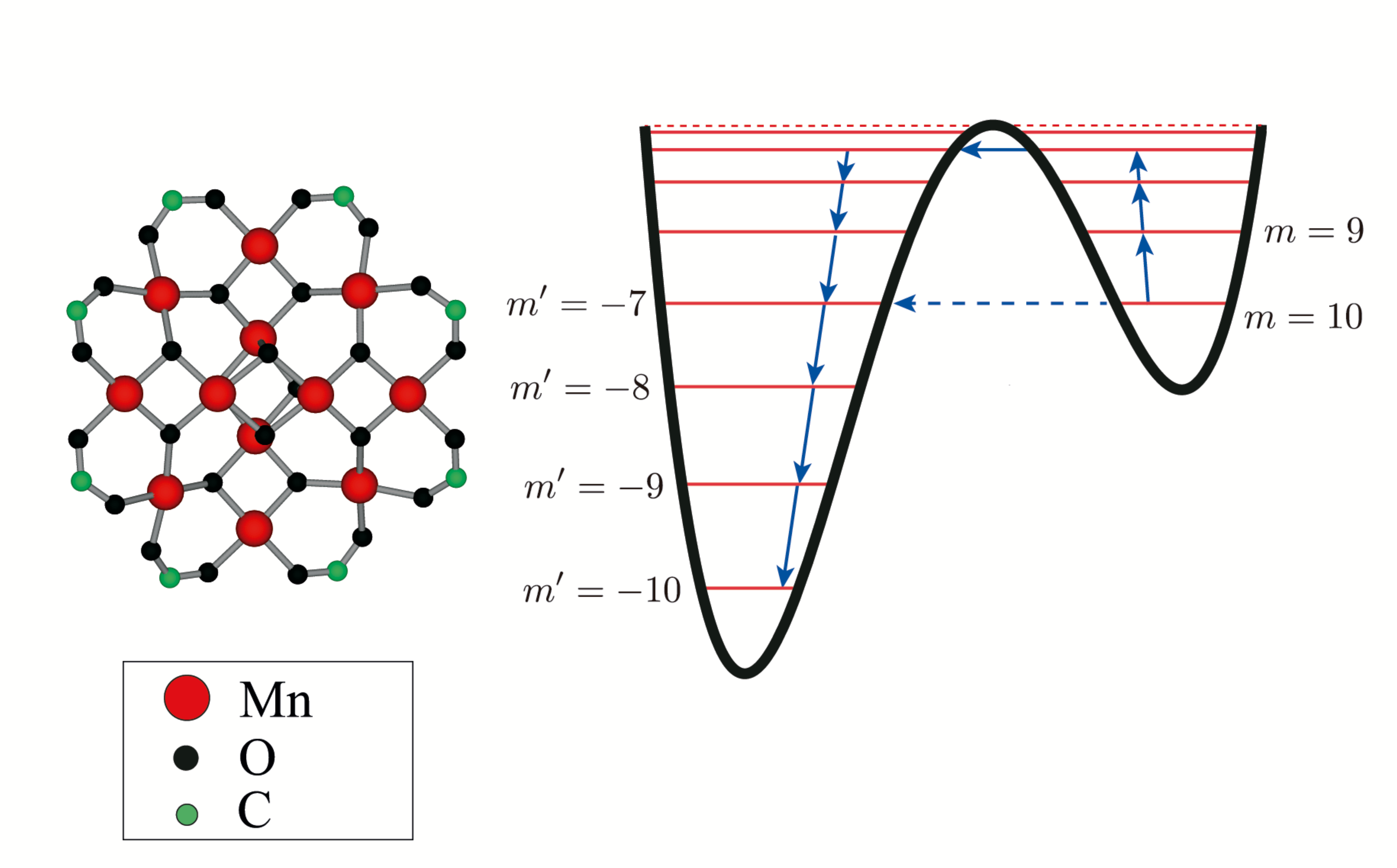}
\caption{Color online: Thermally assisted tunneling between resonant spin levels in a Mn$_{12}$ molecule of spin $S = 10$. The black solid curve shows the dependence of the classical magnetic anisotropy energy on the angle that the magnetic moment forms with the magnetic anisotropy axis. Red lines show energies of spin levels corresponding to magnetic quantum numbers $m$ and $m'$. The dash line illustrates quantum tunneling from $m = 10$ to $m' = -7$.}
\label{fig spin-tunneling}
\end{figure}

Mn$_{12}$ acetate first synthesized by Lis \cite{Lis} is a prototypical spin-10 molecular magnet. It has been studied more than any other molecular magnet. Observation in Mn$_{12}$ acetate of quantum magnetic hysteresis due to resonant spin tunneling  \cite{Friedman-PRL1996} triggered avalanche of papers in this field by chemists and physicists. Spin tunneling in a Mn$_{12}$ molecule is illustrated in Fig. \ref{fig spin-tunneling}. The resonances are achieved on changing the external magnetic field $B$. Their width is due to inhomogeneous dipolar and hyperfine broadening, as well as due to $D$-strains and $g$-strains \cite{Park-PRB2001}. Typically observed widths of the resonances are in the ball park of $1$kOe. The zero-field resonance stands out because it is not subject to $D$-strains and $g$-strains. It also does not depend on whether one works with a single crystal or non-oriented microcrystals. For conventional Mn$_{12}$ acetate the typical width of the zero-field resonance \cite{Friedman-PRB1998} is in the ball park of $300$Oe. Inhomogeneous broadening of spin-tunneling resonances is one obstacle on the way of achieving terahertz lasing and superradiance effects in molecular magnets \cite{chugar-SR,tejada-SR}. 

\begin{figure}[htbp!]
\includegraphics[width=9.0cm,angle=0]{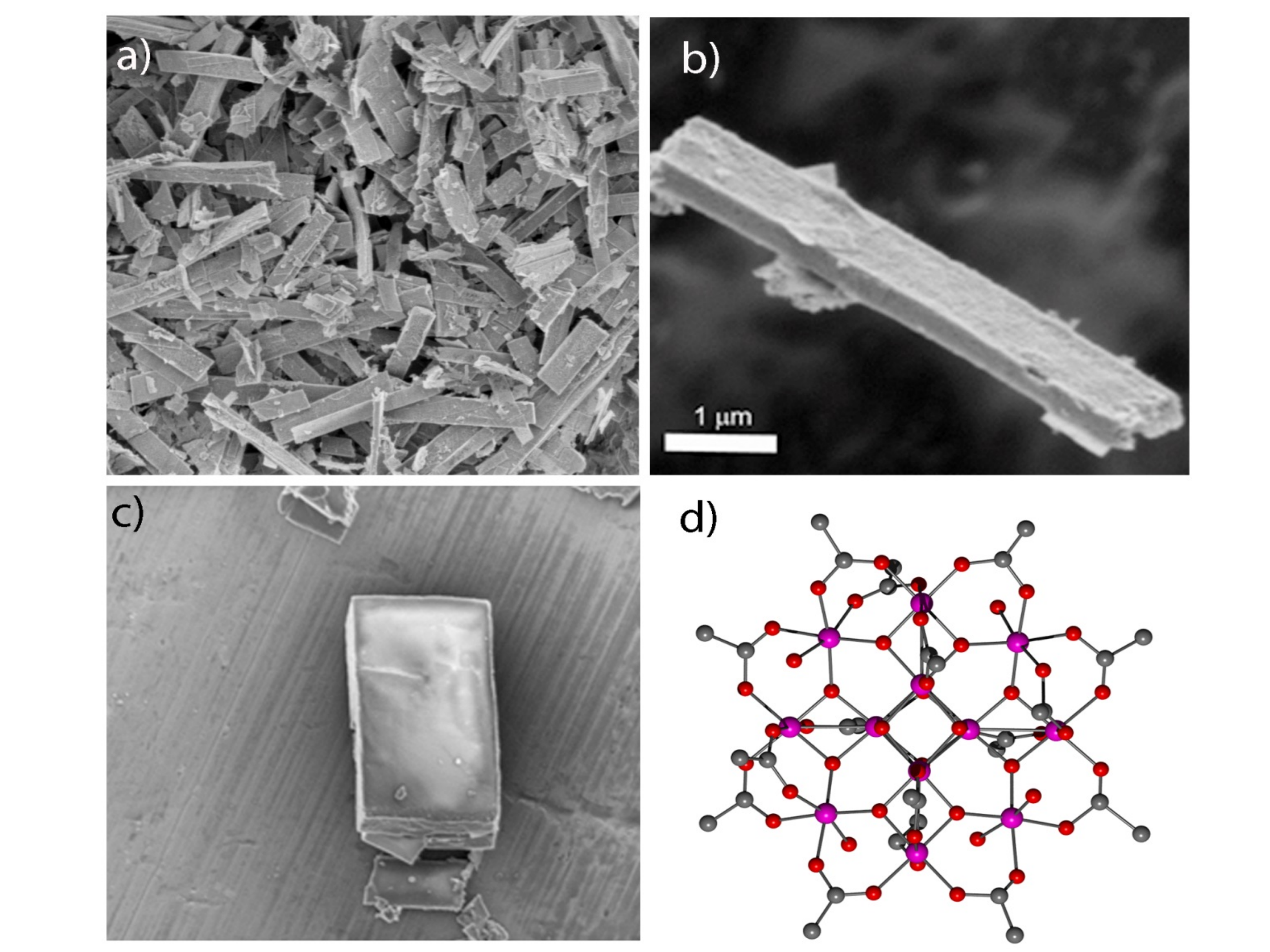}
\caption{Color online: a) FE-SEM images of Mn$_{12}$-acetate ribbons synthesized by reprecipitation in toluene. b) FESEM image of an isolated Mn$_{12}$-Ac ribbon. c) FESEM image of the Mn$_{12}$-Ac triclinic crystal obtained by a slow diffusion of ethanol in a Mn$_{12}$-Ac acetonitrile solution. d) Chemical structure of the Mn$_{12}$-acetate molecule in a triclinic crystal is identical to that in a conventional tetragonal crystal originally synthesized by Lis.}
\label{fig ribbons}
\end{figure}
Recently we reported spin-tunneling maxima in the field derivative of the magnetization of amorphous non-oriented Mn$_{12}$ nanospheres \cite{Lendinez-PRB2015}. This observation challenged the conventional wisdom that quantum resonances can only be observed in single crystals or in systems of oriented microcrystals. The sample was prepared by the method described in Ref. \onlinecite{Imaz-ChemCom2008}. It turned out that only the internal structure of the Mn$_{12}$ molecules was important for the observation of quantum spin tunneling but not their short-range or long-range arrangement in a solid. 

In this paper we report a novel triclinic phase of Mn$_{12}$ acetate with a surprisingly narrow width of the zero-field resonance. In some instances the width as low as $50$Oe has been observed. The paper is structured as follows. Fabrication and characterization of the samples by various techniques are discussed in Section \ref{fabrication}. Measurements of the field and temperature dependence of the magnetization are presented in Section \ref{magnetic}. The results and their possible explanation are discussed in Section \ref{discussion}. 

\section{Fabrication and characterization of triclinic Mn$_{12}$ ribbons}\label{fabrication}
Ribbon-shaped Mn$_{12}$-acetate particles were prepared by reprecipitation \cite{Imaz-ChemCom2008,Lendinez-PRB2015} of Mn$_{12}$-acetate crystals of size $4.9 \pm 1.0 \mu$m that were synthesized as previously described by Lis \cite{Lis}.  This process started with the dissolution of $60$-mg Mn$_{12}$-acetate crystals in $15$mL of acetonitrile. The resulting solution was then filtered to avoid any solid trace and mixed with $30$mL of toluene under continuous stirring, inducing the immediate formation of a brown precipitate. After one hour, the obtained solid was collected by filtration. Field-Emission Scanning Electron Microscopy (FE-SEM) images of this brown solid demonstrated the formation of ribbon-shaped particles, Fig. \ref{fig ribbons}-a,b. The average ribbon size was calculated statistically from FESEM images, measuring the length and the width of $150$ particles of the same sample. The calculated average length was $3.2 \pm 1.9 \mu$m with the median of $2.8 \mu$m, and the average width was $0.8 \pm 0.3 \mu$m with the median $0.8 \mu$m. 

\begin{figure}[htbp!]
\includegraphics[width=9.5cm,angle=0]{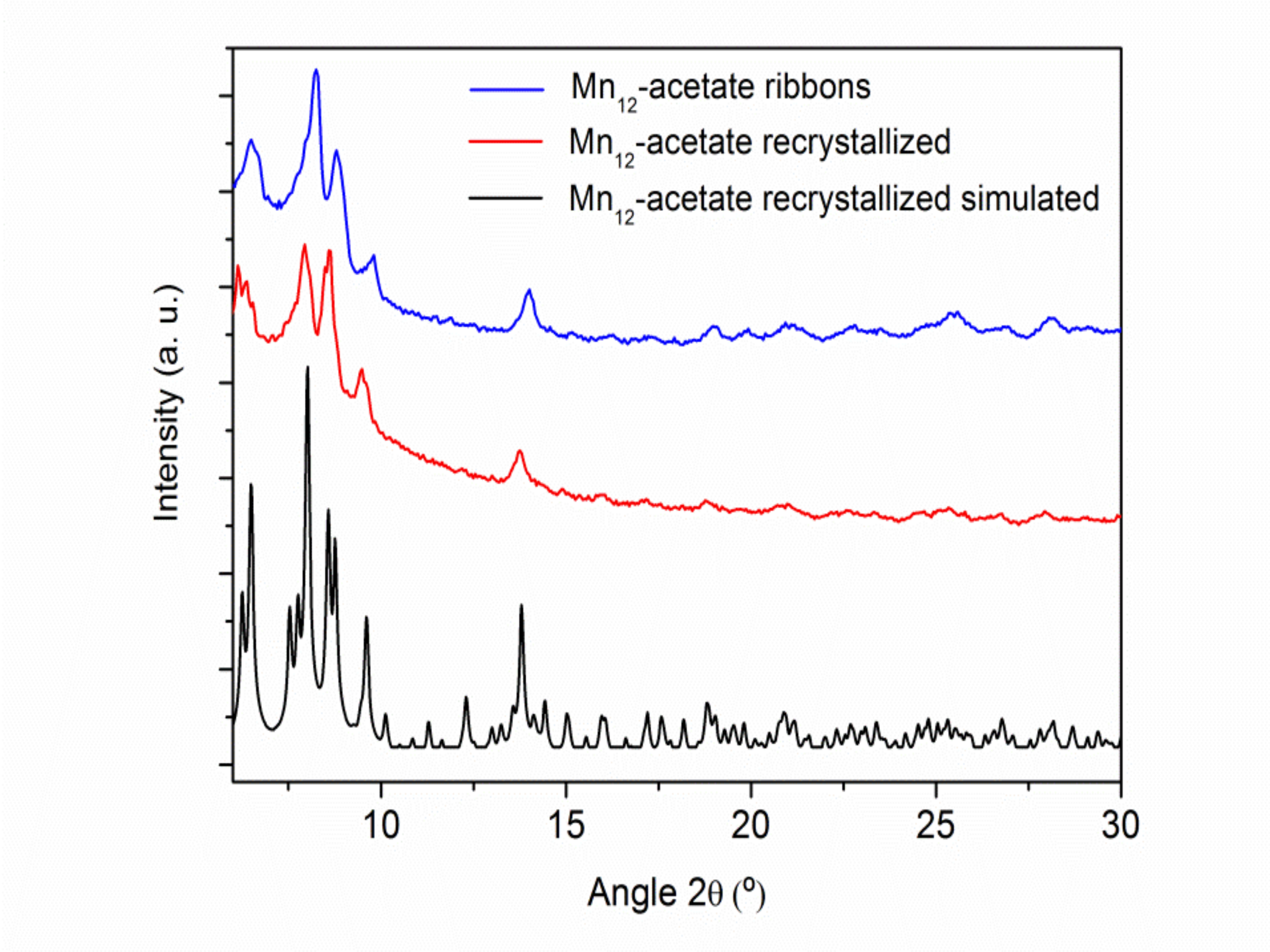}
\caption{Color online: X-ray data from Mn$_{12}$ ribbons (blue), from triclinic crystal grown as described in the text (red), and simulated X-ray pattern from a triclinic crystal (black).}
\label{fig X-ray}
\end{figure}

The X-Ray powder diffraction pattern of the ribbon phase revealed that Mn$_{12}$-acetate ribbons, besides showing amorphous or microcrystalline structure, also did not have the short-range symmetry of the initial Mn$_{12}$-Ac crystal phase,  Fig. \ref{fig X-ray}. In order to identify this new crystalline phase we modified the synthesis of the ribbons to grow single crystals of sufficient size to perform the single-crystal X-ray diffraction experiment. The initial Mn$_{12}$-acetate crystals were dissolved in acetonitrile and the resulting solution was filtered to avoid any solid traces. Then, vapors of toluene were diffused slowly in this solution. After a few days, we observed the formation of rectangular shaped crystals,  Fig. \ref{fig ribbons}-c. 

Due to a small size of the grown crystals, the single crystal X-ray diffraction experiment was performed under Synchrotron radiation in the XALOC beamline at the ALBA synchrotron. The powder pattern simulated from the resolution of the crystalline structure corresponded to the experimental powder diffractogram obtained from the ribbons, confirming their novel short-range crystal structure, Fig \ref{fig X-ray}. 

\begin{figure}[htbp!]
\includegraphics[width=8cm,angle=0]{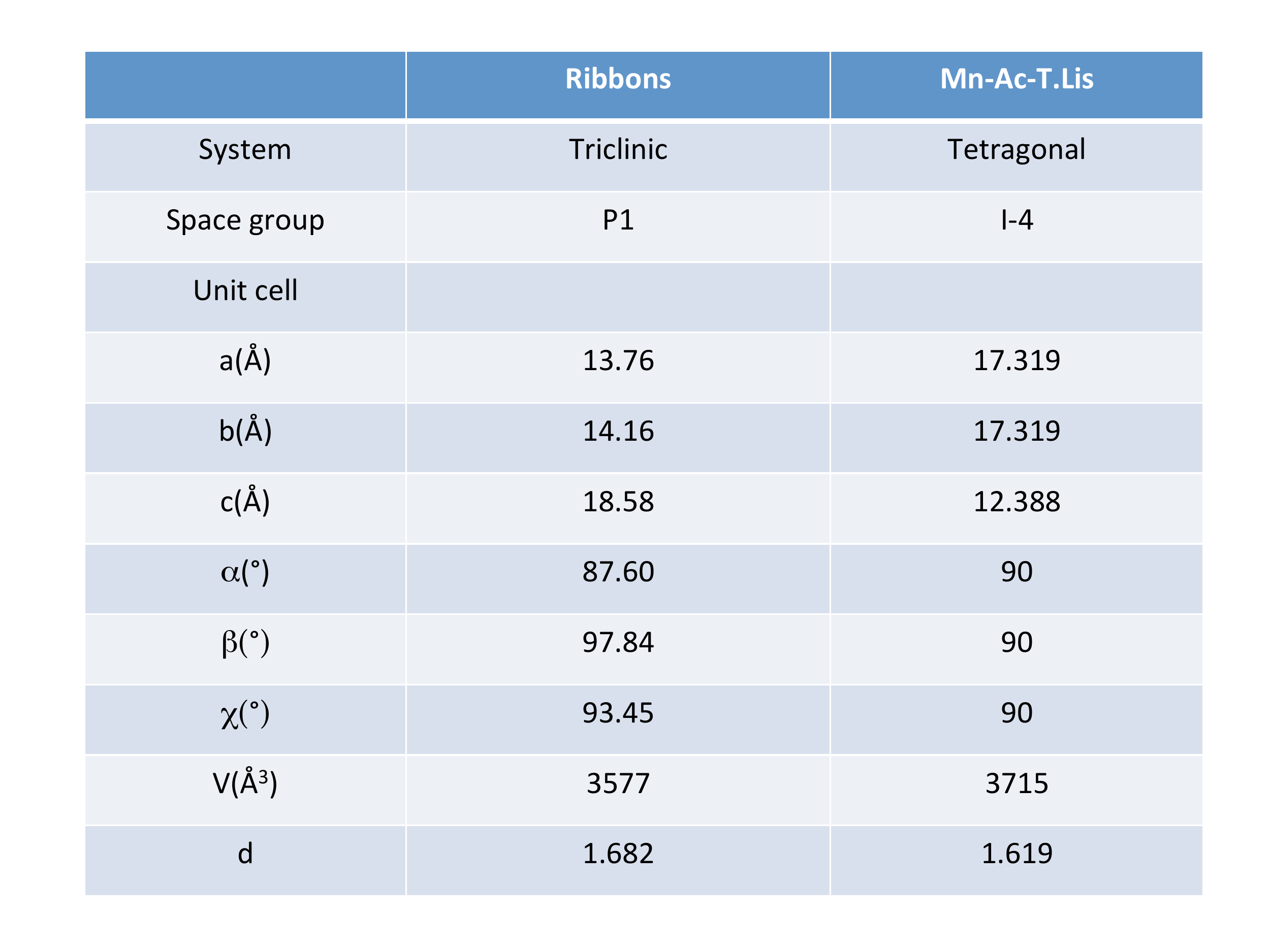}
\caption{Color online: Parameters of triclinic and tetragonal crystals of Mn$_{12}$ acetate.}
\label{fig triclinic}
\end{figure}

The intramolecular structure of the Mn$_{12}$-Ac in the new crystal phase appears identical to the one reported by Lis \cite{Lis}, see Fig. \ref{fig ribbons}-d. The [Mn4O4]8+ cubane unit is held within a non-planar ring of eight Mn3+ ions by eight oxygen atoms. The inorganic core is capped by $16$ acetate molecules. The principal differences between this new phase and the common Mn$_{12}$-Ac phase of Lis resides in the intermolecular packing of the Mn$_{12}$-Ac molecules. In the Lis phase the Mn$_{12}$-Ac molecules crystallize in the tetragonal space group I-4 whereas the new phase crystallizes in the P-1 triclinic space group, Fig. \ref{fig triclinic}.  

\section{Magnetic measurements}\label{magnetic}
\begin{figure}[htbp!]
\includegraphics[width=9.5cm,angle=0]{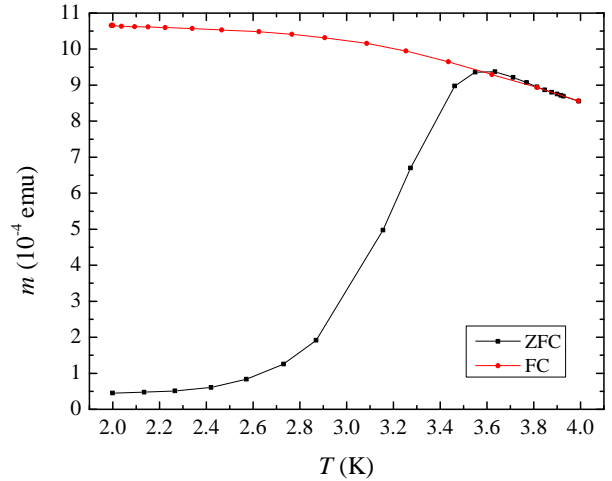}
\caption{Zero-field-cooled (black) and field-cooled (red) magnetization versus temperature for a sample of Mn$_{12}$-acetate ribbons at $H=100$ Oe. A single peak indicates the absence of any second species of Mn$_{12}$ molecules.}
\label{fig ZFC}
\end{figure}

Low-temperature magnetic measurements have been carried out on a compressed powdered sample of Mn$_{12}$ ribbons inside a commercial rf-SQUID Quantum Design magnetometer. Fig. \ref{fig ZFC} shows zero-field-cooled (ZFC) magnetization as a function of temperature. The maximum at about $3.5$K corresponds to the conventional blocking temperature of the Mn$_{12}$-Ac molecule. Below that temperature the molecules hold their magnetic moments while above that temperature they become superparamagnetic. The absence of any secondary maxima in the ZFC curve indicates the presence of only one species of Mn$_{12}$ acetate having a fixed magnetic anisotropy energy of about $65$K. 

\begin{figure}[htbp!]
\includegraphics[width=10cm,angle=0]{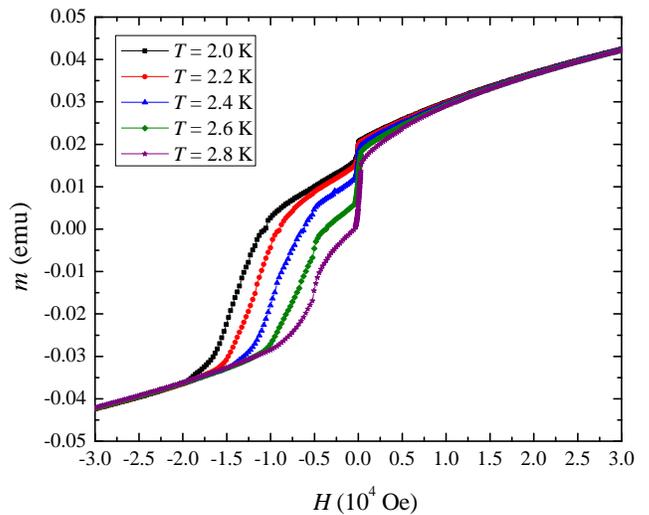}
\caption{Color online: Descending branches of the magnetization curves of Mn$_{12}$ ribbons taken at different temperatures. The data are obtained by first saturating the sample in a positive field and then reducing and reversing the magnetic field. The curves show large jumps at zero field that have not been previously observed in Mn$_{12}$ acetate. }
\label{fig hysteresis}
\end{figure}

Descending branches of the magnetization curves taken at different temperatures are shown in Fig. \ref{fig hysteresis}. The curves have been obtained by first saturating the sample of Mn$_{12}$-acetate ribbons in a positive magnetic field of up to $3$T and then reducing the magnetic field to zero, reversing it, and increasing the field in the opposite direction at a constant rate. Striking feature of these magnetization curves is a very large narrow jump of the magnetization at zero field. In the past such jumps have only been seen in a large magnetic field due to the phenomenon of magnetic deflagration \cite{CCNY-PRL2005,UB-PRL2005}. The latter is equivalent to chemical combustion: Reversal of the magnetic moments of the molecules leads to the release of their Zeeman energy into heat that further enhances magnetic relaxation. However, at $B = 0$ there is no Zeeman energy to burn, so that the deflagration as an explanation to the jumps is ruled out. 

\begin{figure}[htbp!]
\includegraphics[width=10cm,angle=0]{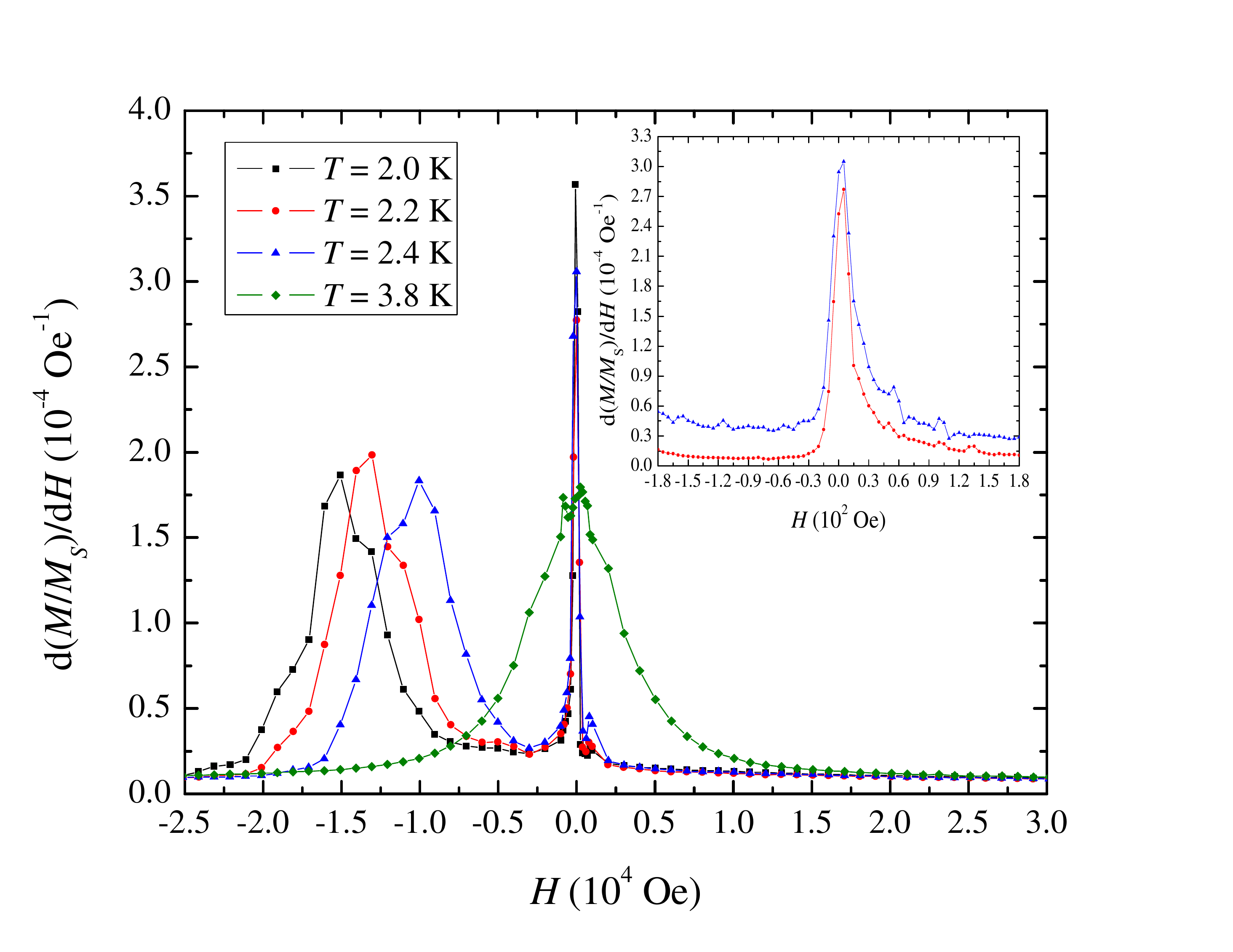}
\vspace{-0.5cm}
\caption{Color online: $d(M/M_{s})/dH$ of Mn$_{12}$ ribbons at different temperatures. Spin-tunneling maxima at zero and first resonant fields are clearly observed. Inset shows low-field structure of the zero-field maximum at 2.2K and 2.4K in steps of 5 Oe.}
\label{fig derivative}
\end{figure}

The derivative of the normalized magnetization, $d(M/M_{s})/dH$, is shown at different temperatures in Fig. \ref{fig derivative}. It clearly indicates four tunneling maxima at the conventional resonant fields of Mn$_{12}$ acetate, separated by about $0.46$T. The presence of such maxima in $d(M/M_{s})/dH$ of non-oriented amorphous particles of a molecular magnet has been explained in Ref. \onlinecite{Lendinez-PRB2015}.  The new feature is an extremely narrow width of the zero-field maximum. In some instances it is below $50$Oe, see inset in Fig. \ref{fig derivative}. Such a narrow spin tunneling resonance has never been observed in any molecular magnet.

\begin{figure}[htbp!]
\includegraphics[width=8.8cm,angle=0]{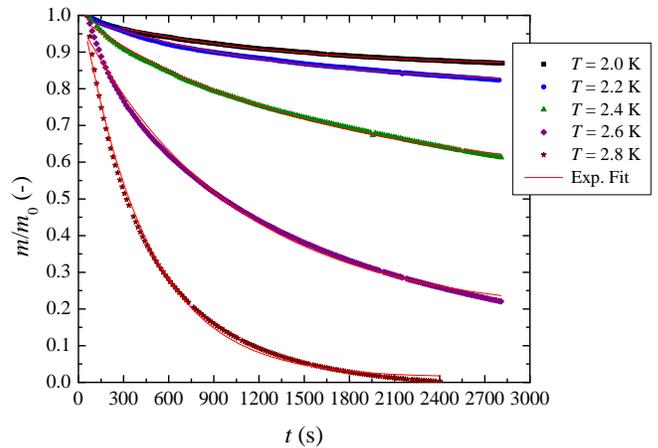}
\vspace{-0.5cm}
\caption{Color online: Exponential time relaxation of the normalized magnetization in a field of -20 Oe at different temperatures --from 2.0 K to 2.8K in increments of 0.2K.}
\label{fig relaxation}
\end{figure}

In accordance with the narrow width of the zero-field resonance one finds an unusually fast magnetic relaxation near zero field, Fig. \ref{fig relaxation}. A single exponential provides a good fit to the time dependence of the magnetic moment of the sample at all temperatures explored (2.0K--2.8K), with mean lifetimes ranging from 438s to 1530s approximately. This means that a single energy barrier contributes to the relaxation. It is determined by the distance from the ground state level $E_0$ to the level $E_m$ that dominates thermally assisted quantum tunneling, see Fig. \ref{fig spin-tunneling}. 

\section{Discussion}\label{discussion}
The jump in the magnetization due to resonant spin tunneling is determined by the incoherent Landau-Zener process \cite{garchu-PRB2002}. The intial, $M_i$, and the final, $M_f$, magnetizations are connected through $M_i/M_f = \exp\{\pi \Delta_m^2\exp[-(E_0 - E_m)/T]/(2v)\}$, where $\Delta_m$ is the tunnel splitting of the level $m$ that dominates thermally assisted quantum tunneling, and $v = 2m\mu_B (dB/dt)$ is the rate at which Zeeman energy is changing due to the field sweep, $\mu_B$ being the Bohr magneton. The double exponential dependence on temperature provides a hint why the size of the magnetization jump in Fig. \ref{fig hysteresis} increases sharply with a relatively small increase of temperature. 

\begin{figure}[htbp!]
\includegraphics[width=9.5cm,angle=0]{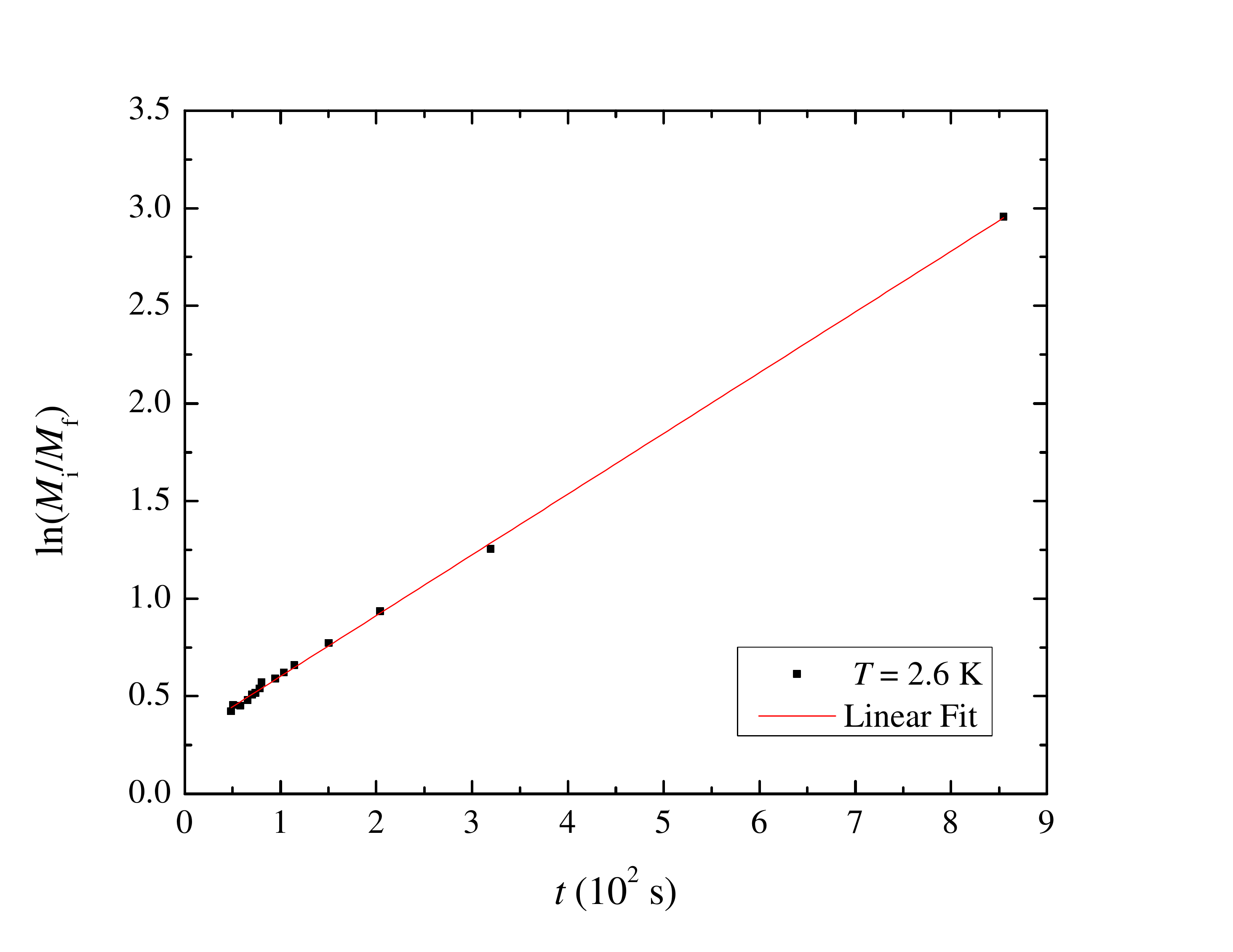}
\vspace{-0.5cm}
\caption{Linear time dependence of $\ln(M_i/M_f)$ during a stepwise field sweep at $T = 2.6$K.}
\label{fig LZ}
\end{figure}
An additional proof of the Landau-Zener dynamics of the magnetization comes from the analysis of the experiment in which the field sweep near zero field on a field reduction from saturation was conducted in equal field steps, with a sizable magnetization change after each step and a little change between the steps. For two consecutive steps one has $(M_{i,n}/M_{f,n})(M_{i,n+1}/M_{f,n+1}) = (M_{i,n}/M_{f,n+1})$ where $M_{i,n+1} = M_{f,n}$ has been used. Writing $(dB/dt)_n$ as $\Delta B/\Delta t_n$, one obtains after $N$ steps
\begin{equation}
\frac{M_{i}}{M_{f}} = e^{\frac{\pi \Delta_m^2 e^{-(E_0-E_m)/T}}{4m \mu_B  \Delta B}t}, \hspace{0.5cm} t = \Delta t_1 + \Delta t_2 + .... + \Delta t_N
\end{equation}
resulting in  $\ln\left({M_{i}}/{M_{f}}\right) \propto t$. The linear time dependence of $\ln\left({M_{i}}/{M_{f}}\right)$ observed in experiment is illustrated by Fig. \ref{fig LZ}.

\begin{figure}[htbp!]
\includegraphics[width=9cm,angle=0]{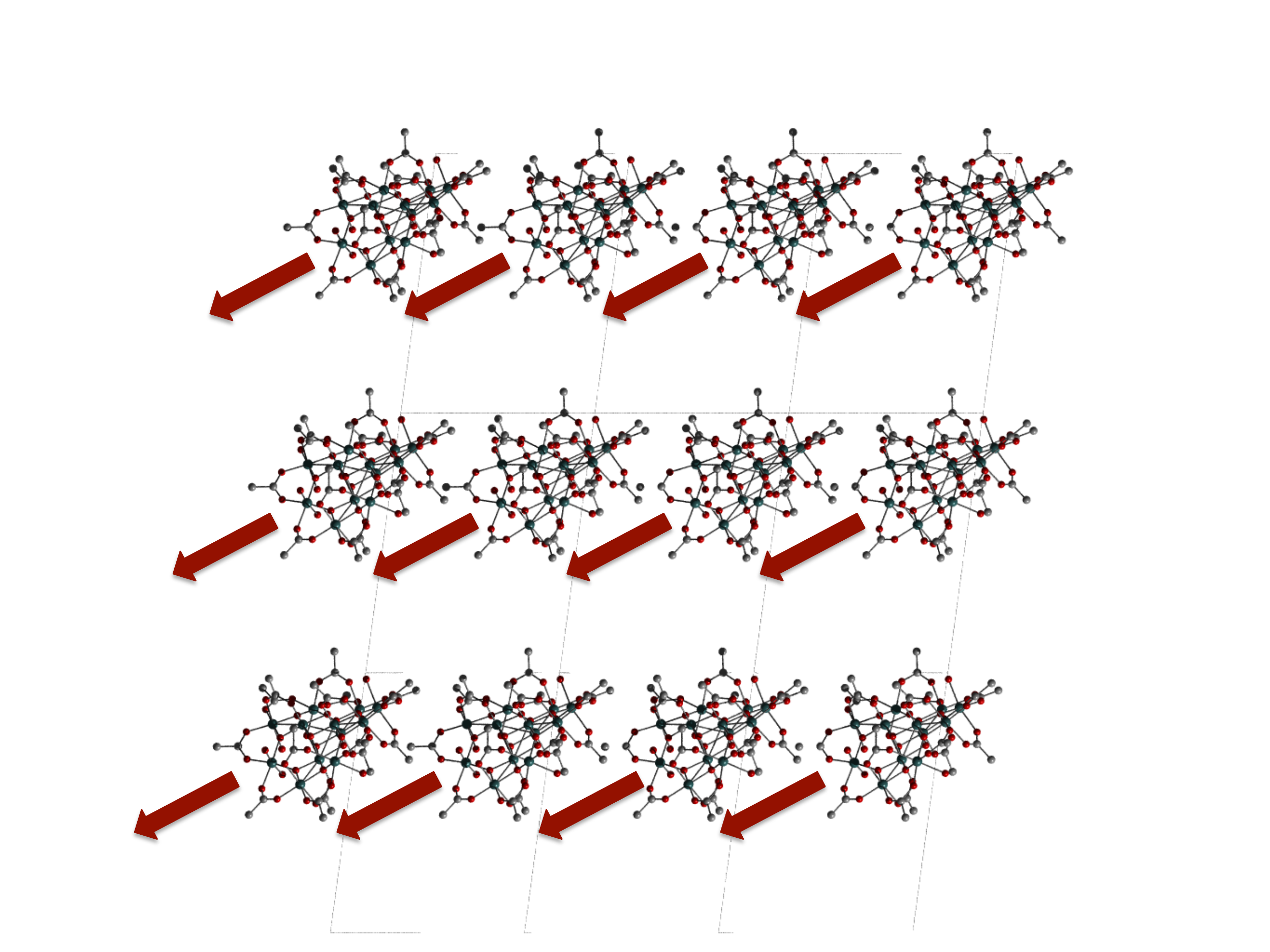}
\caption{Color online: Triclinic short-range crystal order in the arrangement of Mn$_{12}$-acetate molecules in the ribbons. Arrows (conceptual) show directions of the magnetic moments of the molecules. Unlike in a conventional tetragonal crystal, the moments do not form chains along a single magnetic anisotropy axis. }
\label{fig arrangement}
\end{figure}

The width of the zero-field resonance is determined by the inhomogeneous broadening due to dipole-dipole interaction between magnetic moments of the Mn$_{12}$ molecules and due to hyperfine interactions that are rather strong in Mn$_{12}$. Strong dipolar broadening in conventional tetragonal crystals of Mn$_{12}$ acetate is due to the tendency of the molecules to form ordered chains along the tetragonal C-axis. This tendency should be greatly diminished in the triclinic structure. In the new crystal phase of Mn$_{12}$ acetate reported here the magnetic moments of the molecules do not form chains, see Fig. \ref{fig arrangement}. Still it appears unlikely that the dipolar broadening in the ribbons would fall below $50$Oe. 

Even if the dipolar broadening of the resonance was very low, the question would remain about the absence of the hyperfine broadening, which should be of the order of a few hundred Oe. One possible explanation to the data is provided by recent theoretical work that shows that the magnetization reversal in a molecular magnet may occur inside a front of quantum tunneling moving through the sample. This may happen either due to the formation of a domain wall with a zero dipolar bias inside the wall \cite{garchu-PRB2008} or due to the formation of a self-organized front of quantum tunneling with zero bias between tunneling spin levels regardless of the nature of inhomogeneous broadening \cite{garchu-PRL2009}. The latter effect presents the most interesting possibility. It is similar to the optical laser effect: The dipolar field in a system of magnetic dipoles self-organizes to provide the fastest relaxation to the minimum energy state. It has been demonstrated \cite{garchu-PRL2009} that self-organization of dipolar field can provide the resonant condition in the presence of a very significant (up to 25\%) broadening of the energy of spin levels. Further studies will reveal if self-organization is the reason for the observed narrow zero-field resonance. If it is, it may open the way of lasing with molecular magnets in the terahertz frequency range.\\

\section{Acknowledgements}

The work at the University of Barcelona has been supported by the Spanish Government Project No.  MAT2011-23698.  I.I. and J.E. thank the MINECO for the Ram\'{o}n y Cajal contract and the FPI fellowship, respectively. The work of E.M.C. at Lehman College has been supported by the U.S. National Science Foundation through grant No. DMR-1161571.

\end{document}